\documentclass{midl} 

\usepackage{mwe} 
\jmlrvolume{-- Under Review}
\jmlryear{2022}
\jmlrworkshop{Full Paper -- MIDL 2022 submission}
\editors{Under Review for MIDL 2022}

\title[Cross-Domain Federated Learning in Medical Imaging]{Cross-Domain Federated Learning in Medical Imaging}

\midlauthor{\Name{Vishwa S Parekh \nametag{$^{1,2}$}} \Email{vishwaparekh@jhu.edu}\\
\Name{Shuhao Lai \nametag{$^{1}$}} \Email{slai16@jhu.edu}\\
\Name{Vladimir Braverman \nametag{$^{1}$}} \Email{vova@cs.jhu.edu}\\
\Name{Jeff Leal \nametag{$^{2}$}} \Email{jleal1@jhmi.edu}\\
\Name{Steven Rowe \nametag{$^{2}$}} \Email{srowe8@jhmi.edu}\\
\Name{Jay J. Pillai \nametag{$^{2}$}} \Email{jpillai1@jhmi.edu}\\
\Name{Michael A Jacobs \nametag{$^{2,3}$}} \Email{mikej@mri.jhu.edu}\\
\addr $^{1}$ Department of Computer Science, The Johns Hopkins University, Baltimore, MD 21218 \\
\addr $^{2}$  The Russell H. Morgan Department of Radiology and Radiological Sciences, The Johns Hopkins University School of Medicine, Baltimore, MD 21205, USA \\
\addr $^{3}$ Sidney Kimmel Comprehensive Center, The Johns Hopkins University School of Medicine, Baltimore, MD 21205, USA
}

\begin{document}

\maketitle

\begin{abstract}
Federated learning is increasingly being explored in the field of medical imaging to train deep learning models on large scale datasets distributed across different data centers while preserving privacy by avoiding the need to transfer sensitive patient information. In this manuscript, we explore federated learning in a multi-domain, multi-task setting wherein different participating nodes may contain datasets sourced from different domains and are trained to solve different tasks. We evaluated cross-domain federated learning for the tasks of object detection and segmentation across two different experimental settings: multi-modal and multi-organ. The result from our experiments on cross-domain federated learning framework were very encouraging with an overlap similarity of 0.79 for organ localization and 0.65 for lesion segmentation. Our results demonstrate the potential of federated learning in developing multi-domain, multi-task deep learning models without sharing data from different domains.
\end{abstract}


\section{Introduction}
Deep learning techniques are being developed and applied across various applications in the field of Radiology \cite{suzuki2017overview,mcbee2018deep,parekh2019deep}. There are various intrinsic similarities across different deep learning models being evaluated in radiological applications. For example, segmentation models developed for multiparametric (mp) MRI would learn a similar tissue signature for segmenting different lesions irrespective of the underlying application being evaluated (e.g. brain , breast , or prostate MRI). Similarly, an object detection model built for localizing organs in a whole-body (WB) MRI would would share architectural (anatomical) similarities with object detection models built for localizing organs in a different modality like WB-PET or WB-CT. As a result, deep learning models being built for different tasks across diverse domains would significantly benefit from sharing knowledge or collaborative training. 

The applicability of deep learning models in a cross-domain setting for medical imaging has previously been explore across many applications \cite{samala2018cross,hadad2017classification,tomczak2020multi,parekh2020multitask}. However, current applications in training cross-domain deep learning models require centralized datasets and have several challenges. For example, a medical imaging center may not be equipped with PET scanners and sourcing PET data from a different medical imaging center could be difficult owing to patient privacy concerns. Similarly, different medical imaging centers could be developing deep learning models for related tasks; suppose one center is training a breast tumor segmentation model from mpMRI while another center is training a normal tissue segmentation model in breast mpMRI for breast density estimation. Cross-domain collaborative, decentralized learning in these scenarios could have numerous benefits. First, different centers lacking specialized equipment (e.g. imaging  equipment) could train models to learn PET scan signals without having to source PET scans from other institutions. Second, annotating datasets is expensive and time consuming. Different institutions could potentially benefit from each others' annotations without explicitly sharing them. The third benefit could be observed in terms of computational and space efficiency of training and storing comparatively fewer deep learning models.

Collaborative, decentralized training could be achieved using federated learning \cite{mcmahan2017communicationefficient, changDistributed}. Federated learning is a machine learning paradigm that aims to train a machine learning algorithm (e.g. deep neural network) on multiple datasets stored across decentralized nodes. The local datasets stored on different nodes are not shared, therefore, preserving privacy. The global machine learning model is trained by training several local machine learning models across different nodes that exchange model parameters (e.g. weights and biases of a deep neural network) at a certain frequency which are then aggregated and shared back to the local nodes. Federated learning is increasingly being explored in the field of medical imaging to train deep learning models on large-scale datasets distributed across different data centers while preserving privacy \cite{changDistributed, ng2021federated, sarma2021federated,shen2021multitask,qayyum2021collaborative}. 

In this work, we explore the potential of cross-domain federated learning across two different experimental setups. The first experiment involves learning an organ localization model across two imaging modalities, PET and CT. The second experiment involves learning a lesion segmentation model across two organs, brain and breast, using mpMRI. The experiments are detailed in Section 2, followed by results in Section 3 and Discussion in Section 4. 

\section{Experiments}
\subsection{Experiment 1: Multi-modal organ localization}
\subsubsection{Clinical Data}
All studies were performed in accordance with the institutional guidelines for clinical research under a protocol approved by our Institutional Review Board (IRB) and all HIPAA agreements were followed for this retrospective study. The clinical data consisted of fifty patients. Patients were imaged on a Biograph mCT 128-slice PET/CT scanner (Siemens Healthineers). For the 18F-DCFPyL scans, patients were intravenously injected with no more than 333 MBq (9 mCi) of radiotracer approximately 60 min before image acquisition.  The field of view was vertex to mid thigh for 18F-DCFPyL. The WB PET-CT images were evaluated in this study for localization of kidneys. 
\subsubsection{Deep learning model}
The deep learning model used for localization of kidneys was a U-Net model \cite{ronneberger2015u}. The encoder, decoder, and the bridge sections of the U-Net consisted of five, one, and five convolutional blocks, respectively. Each covolutional block comprised of two convolutional layers with ReLU activation, followed by batch normalization. A max pooling layer (window size=2x2) preceded the encoder convolutional blocks, while the decoder convolutional blocks were preceded by an unpooling layer, followed by concatenation with the layer at the same level in the encoder section, as shown in Figure 1.  
\begin{figure}[htbp]
\floatconts
  {fig:example1}
  {\caption{Illustration of the U-Net model used for segmentation and localization of different regions of interest across all the experiments}}
  {\includegraphics[width=1\linewidth]{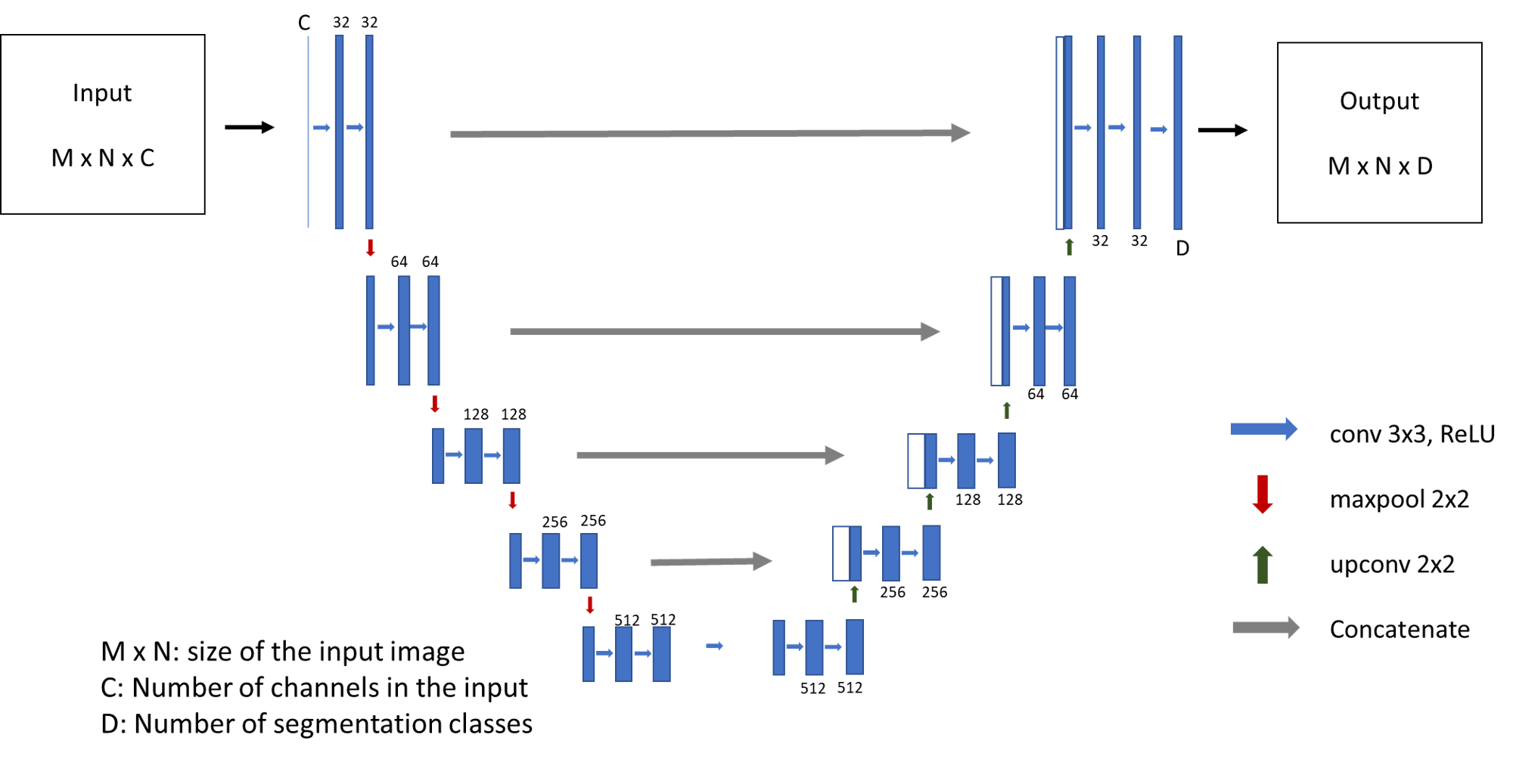}}
\end{figure}
\subsubsection{Experimental Setup}
The federated learning set up for this experiment consists of a central server and two client nodes as shown in Figure 2. The fifty patients were randomly split into train-test subsets with 35 patients used for training and 15 patients used for testing. The training set was divided across two nodes, with the first node comprising of CT images and the second node comprising of PET images. 

The global U-Net model for localization of kidneys across PET and CT modalities was trained by the central server using the following steps:
\begin{itemize}
\item Step 1: The central server initializes the global U-Net model. 
\item Step 2: Repeat Steps 3 to 6 N times, where N corresponds to the number of communication iterations. 
\item Step 3: The central server communicates the global U-Net model to each node
\item Step 4: Each node locally trains the model on their respective local datasets for one epoch. 
\item Step 5: The updated model parameters are communicated back to the central server from each node.
\item Step 6: The central server updates the global U-Net model by aggregating the updated model parameters from each node using the Federated Averaging (FedAvg) algorithm.
\end{itemize}
\begin{figure}[htbp]
\floatconts
  {fig:example2}
  {\caption{Illustration of the U-Net model used for segmentation and localization of different regions of interest across all the experiments}}
  {\includegraphics[width=1\linewidth]{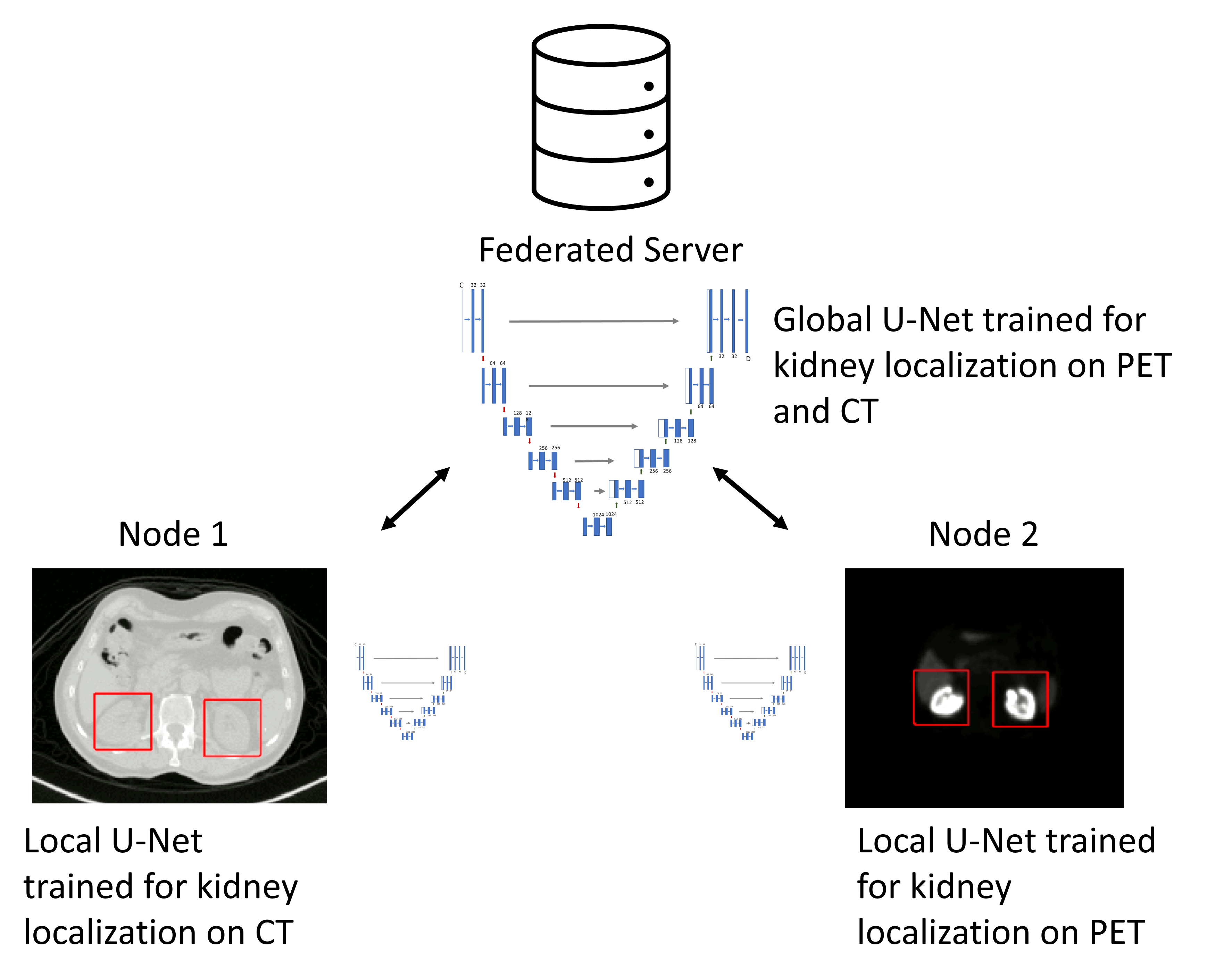}}
\end{figure}
The local U-Net models were trained with a batch size of four. The output activation function was set to Sigmoid activation. The loss function was set to binary cross-entropy and the Adam optimizer with a learning rate of 0.0002 was used for this experiment. The performance of the global U-Net model on the test dataset was evaluated using a modified overlap similarity metric. The metric was computed by dividing the intersection between the organ boundary and the bounding box by the organ boundary to determine if the bounding box was able to bound the organ entirely or not.

\subsection{Experiment 2: Multi-organ tumor segmentation}
\subsubsection{Clinical Data}

\paragraph{Breast mpMRI}
Fifty patients (25 malignant and 25 benign lesions were included in this experiment. Patients were scanned on a 3T MRI system (3T Achieva, Philips Medical Systems, Best, The Netherlands) using a bilateral, dedicated phased array breast coil (InVivo, Orlando, FL) with the patient in the prone position. The complete acquisition details have been described in \cite{parekh2017integrated}. The following imaging sequences were acquired for each patient. 
\begin{itemize}
    \item T1-weighted MRI
    \item Fat-suppressed T2-weighted MRI
    \item Diffusion Weighted Images (DWI)/Apparent Diffusion Coefficient (ADC) map acquired at b values of  b0 and b600
    \item High temporal resolution dynamic contrast enhanced (DCE) MRI (One pre- and fourteen post-contrast images).
    \item High spatial resolution DCE-MRI (One pre- and one post-contrast image). 
\end{itemize}

\paragraph{Brain mpMRI}
The brain mpMRI comprised of 50 patients randomly selected from the brain tumor segmentation (BRATS 2017) challenge dataset. Of the 50 patients, 25 patients had High Grade Glioma and 25 patients had Low Grade Glioma. The mpMRI consisted of pre and post 3D T1-weighted, 2D T2-weighted, and T2-weighted FLAIR images. Complete MRI acquisition parameters can be found in \cite{menze2014multimodal}

\subsubsection{Deep learning model}
The U-Net architecture, shown in Figure 1 used for the localization of kidneys was also used for segmentation of brain and breast lesions.

\subsubsection{Experimental Setup}
The federated learning set up for this experiment consists of a central server and two client nodes similar to the one shown in Figure 2. The fifty patients from each group were randomly split into train-test subsets with 35 patients used for training and 15 patients used for testing. The training set was divided across two nodes, with the first node comprising of breast mpMRI images and the second node comprising of brain mpMRI images. 

The imaging sequences of T1-weighted pre- and post-contrast enhanced MRI, and T2-weighted images were used to train the U-Net model for tumor segmentation. The global U-Net model for segmentation of both breast and brain lesions were trained using the steps outlined in Section 2.1.3

The local U-Net models were trained with a batch size of four. The loss function was set to binary cross-entropy and the Adam optimizer with a learning rate of 0.0002 was used for this experiment. The performance of the global U-Net model on the test dataset was evaluated using the dice-similarity metric. 

\section{Results}
The cross-domain federated learning demonstrated encouraging results across both the experiments as shown in figures 3 and 4. The test set for kidney localization consisted of a total of thirty cases with fifteen cases each for PET and CT scans. The average overlap similarity for kidney localization was 0.79 across both modalities. For the lesion segmentation experiment, the test set consisted of fifteen brain mpMRI and fifteen breast mpMRI dataset. The average dice similarity for lesion segmentation across brain and breast mpMRI was 0.65 and the model failed to segment the lesion on three cases (one brain and two breast cases). 
\begin{figure}[htbp]
\floatconts
  {fig:example3}
  {\caption{Illustration of the performance of the U-Net model for localization of kidneys across PET and CT}}
  {\includegraphics[width=0.5\linewidth]{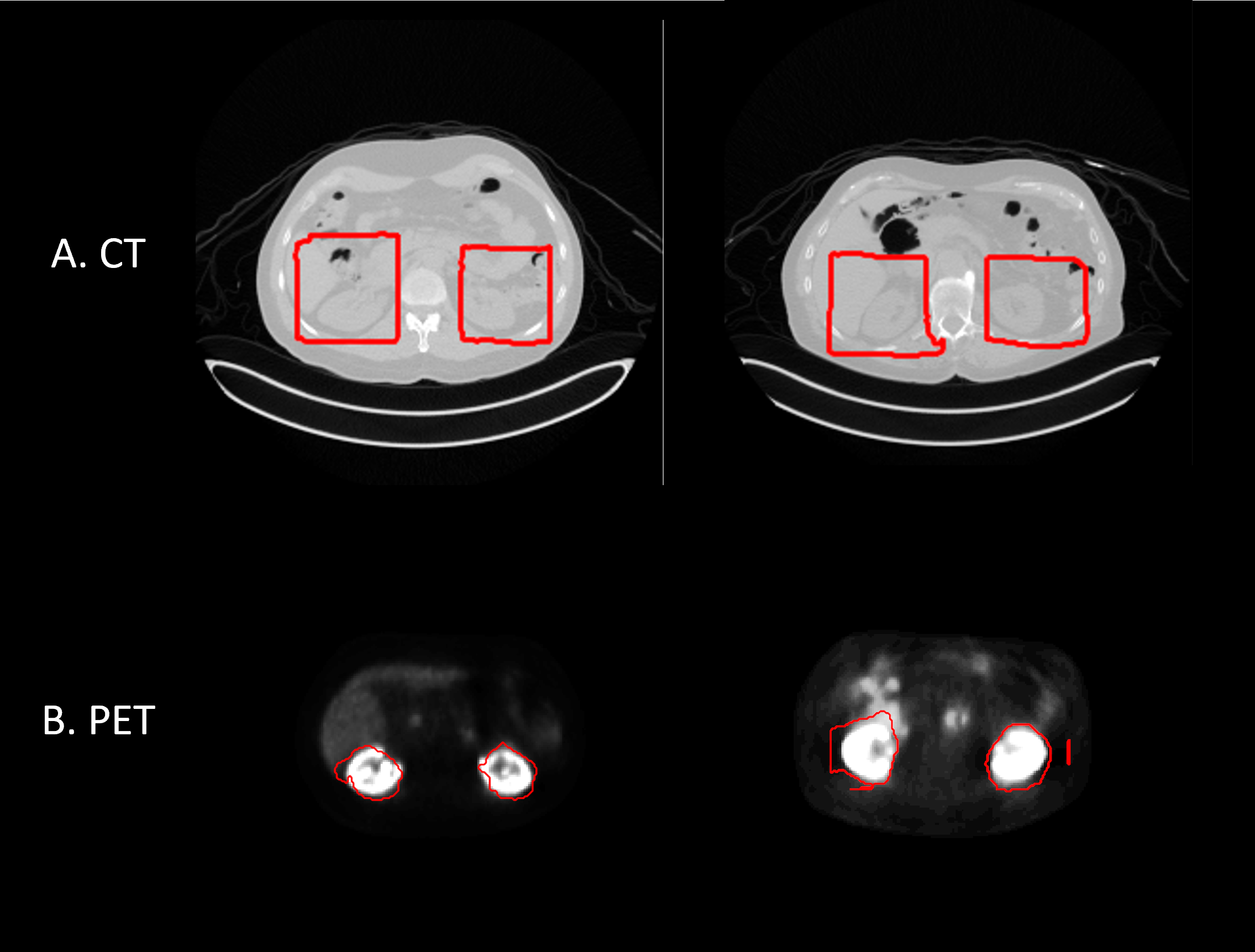}}
\end{figure}
\begin{figure}[htbp]
\floatconts
  {fig:example4}
  {\caption{Illustration of the performance of the U-Net model for lesion segmentation across brain and breast}}
  {\includegraphics[width=0.5\linewidth]{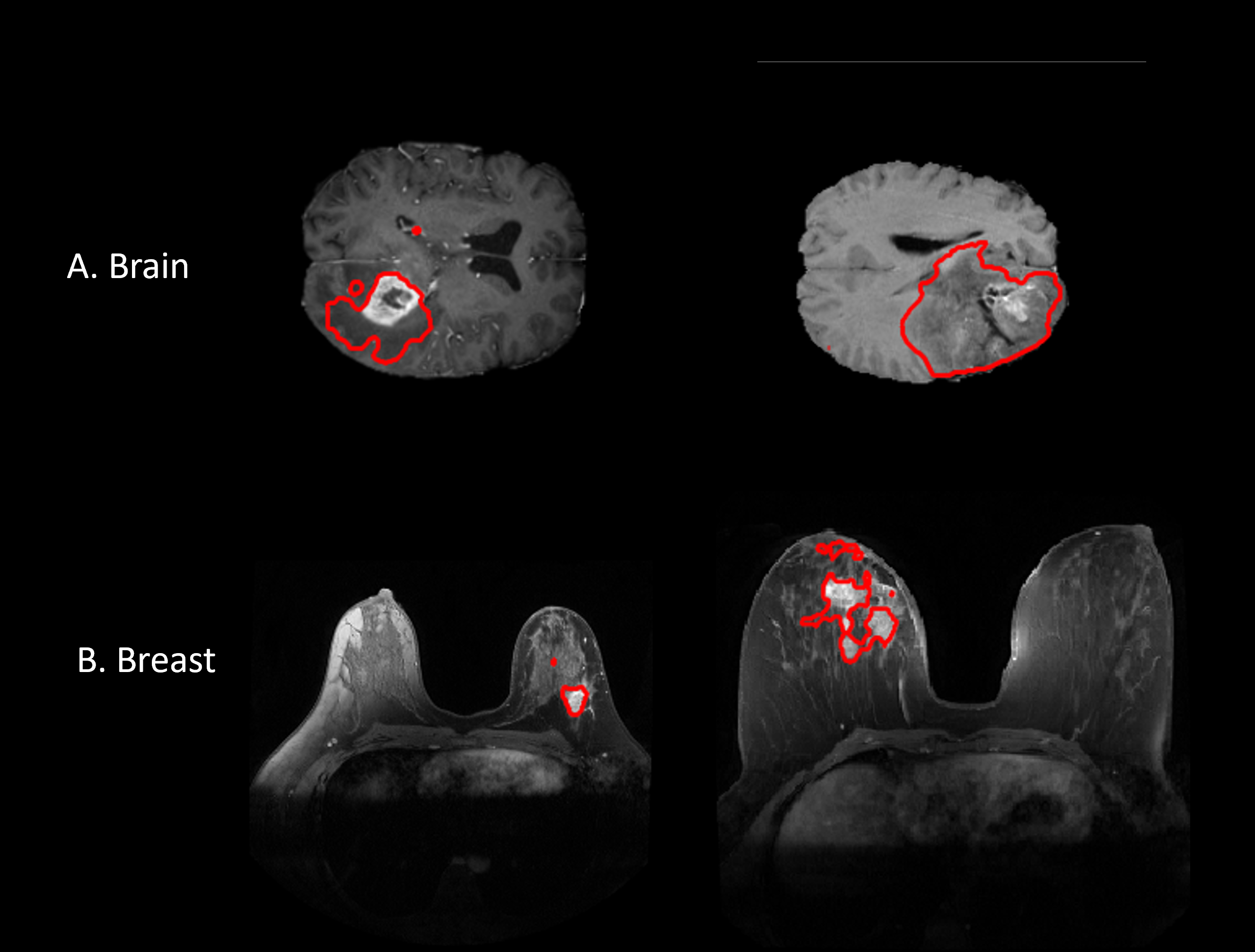}}
\end{figure}
\section{Discussion}
We demonstrated the feasibility of training cross-domain federated learning models with encouraging results. Our results assert the possibility of training a general AI framework that can leverage the unique domain and tasks provided by participating institutions. Federated learning has previously been successfully applied for single medical imaging segmentation tasks \cite{li2019privacy,changDistributed, ng2021federated, sarma2021federated,shen2021multitask,qayyum2021collaborative}. However, our work broadens the possibilities for application of federated learning in the field of medical imaging, especially when participating institutions lack specialized equipment (e.g. PET/CT scanners).

Deep learning techniques are increasingly being used in radiological applications. However, an increase in the number of applications where deep learning could be applied to radiology makes the maintenance and deployment of different deep learning models to radiology computationally very expensive. Our approach offers a solution by collaboratively building multi-task, multi-domain deep learning models.

Our study has certain limitations. This is a preliminary study with experiments performed on small datasets to establish a proof of concept. The use of small datasets is also observed in our results, which are encouraging but not comparable to state-of-art established in the literature. The federated learning model was able to identify the correct region of interest on most of the patients across both datasets, with an exception of three cases. However, the model failed to localize the entire region of interest in many cases resulting in low dice similarity metrics. In the future, we plan to optimize our models and expand our experiments on large-scale datasets. In conclusion, this work initiates a first step towards a cross-domain federated learning setup and explores its potential in building a collaborative multi-domain network of participants sharing heterogeneous datasets.

\midlacknowledgments{National Institutes of Health (NIH) grant numbers: 5P30CA006973 (Imaging Response Assessment Team-IRAT), U01CA140204. Defense Advanced Research Projects Agency DARPA-PA-20-02-11-ShELL-FP-007.}

\bibliography{midl-samplebibliography}

\end{document}